\documentclass[journal]{IEEEtran}
\usepackage{cite}
  \usepackage[pdftex]{graphicx}
\usepackage[caption=false,font=footnotesize]{subfig}
\usepackage{stfloats} \fnbelowfloat
\usepackage{url}
\begin{document}
\title{A Scenario for a Future European Shipboard Railgun}
\author{\IEEEauthorblockN{Stephan Hundertmark and Daniel Lancelle}
}

\maketitle
\begin{abstract}
Railguns can convert large quantities of electrical energy into kinetic energy of the projectile.
This was demonstrated by the 33\,MJ muzzle energy shot performed in 2010 in the framework
of the Office of Naval Research (ONR) electromagnetic railgun program.
Since then, railguns are a prime candidate for future long range artillery systems. In this
scenario, a heavy projectile (several kilograms) is accelerated to
approx. 2.5\,km/s muzzle velocity. While the primary interest for such
a hypersonic projectile 
is the bombardment of targets being hundreds of kilometers away, they can also be used to counter
airplane attacks or in other direct fire scenarios. In these cases, the large initial velocity 
significantly reduces the time to impact the target. In this study we investigate a scenario, 
where a future shipboard railgun installation delivers the same kinetic energy to a target as 
the explosive round of a contemporary European ship artillery system. At the same time the railgun
outperforms the current artillery systems in range. For this scenario a first draft for the parameters
of a railgun system were derived. For the flight-path of the projectile, trajectories for
different launch angles were simulated and the aero-thermodynamic heating was estimated using
engineering-tools developed within the German Aerospace Center (DLR). This enables the
assessment of the feasibility of the different strike scenarios, as well as the
identification of the limits of the technology. It is envisioned that this baseline design can be used as
a helpful starting point for discussions of a possible electrical weaponization of future European
warships.
\end{abstract}

\section{Introduction}
One of the main selling points for a railgun is, that it can convert
large quantities of electrical energy into kinetic energy of a
projectile. At the same time it was repeatedly demonstrated that
muzzle velocities of 2.5\,km/s can be realized. These two capabilities
uniquely qualify railguns as a candidate for a long range artillery
system. One of the military platforms, where such a gun could be
deployed on, is a larger ship, i.e. a future all-electric frigate or 
destroyer. In an all-electric ship, the electrical engine needs to be
able to accelerate these large ships to velocities of 20\,kn to
30\,kn, thus requiring an electrical power of somewhere between 30\,MW
to 100\,MW. As most of the time, the ship will not need all of its
installed power for the drive system, it is natural to equip such a
vessel with electrical weapons. The railgun and the high energy laser
are prime candidates for such new, electrical weapon systems. When
looking at these two systems, the railgun is closely related to the
traditional artillery guns being mounted on the current ships. It launches a
projectile on a ballistic trajectory, with the only difference to use
a magnetic field instead of gun powder as propellant. As a railgun
uses constant acceleration over the barrel length it can achieve
higher muzzle velocities than conventional guns of the same length.
This higher velocity, in combination with a hypersonic projectile
design, translates into a greatly extended range. From the
capabilities point of view, a railgun can do all what a conventional gun
can do, but better. The navies of the European Union member states
have a combined fleet of about 110 frigates and destroyers currently
in service \cite{wiki_europe_ships}. In the future, these will have to be gradually
replaced by modern vessels with an electric drive. Even so the total number of
ships might shrink due to budget constraints, there is clearly a
large market for railgun equipped ships. In this study, it is
attempted to develop the key parameters of a railgun system that is
needed to match and exceed the capabilities of current shipboard
artillery. In addition flight behavior of a first draft for a
hypersonic projectile is evaluated using standard software for
preliminary missile design.

\section{Artillery Capabilities of Current Ships}
In France and Germany, there is no clear distinction in name between
frigates and destroyers. Instead both types are referred to as frigates. 
The frigates of both navies are mainly equipped with two calibers for the
main gun. Most of the French vessels have a 100\,mm caliber cannon,
named "modele 68" or a variant of it mounted \cite{wiki_100mm}.
The German frigates are equipped with the smaller 76\,mm caliber gun 
from Oto-Melara \cite{wiki_otobreda}. Table \ref{current_art} lists
the most important parameters for these two weapons. The ratio of
the projectile to total round mass is about 50\,\% to 56\,\%. The
standard ammunition for these guns uses an explosive warhead.
Therefore the amount of carried explosive determines the amount of 
energy delivered to the target. As
an estimate for this energy level one can use the energy content of
TNT and scale it by the weight of the bursting-charge. For the
76\,mm gun, the energy released at the target is about 2\,MJ, while
the 100\,mm gun delivers 4\,MJ. Of course these two numbers are only a
superficial criterion, as another important parameter is the accuracy
with which a target can be hit. Even so there is not an a priori reason,
as to why railguns could not be used to launch explosive rounds, there
is a certain charm in the idea to use the large velocity delivered by
railguns to cause the destruction at the target by kinetic energy only. 
This has the advantage, that it eliminates the need for the costly chain of
production, storage, delivery and handling of explosives in addition
to reduce the vulnerability of the vessel. For a comparable
effect to the existing armament of the current naval vessels, the amount 
of kinetic energy with which a railgun projectile needs to impact is of 
the order of 2\,MJ to 4\,MJ. Using an explosive warhead would allow to
strongly reduce the impact and therefore the muzzle velocity. The artillery 
range capabilities of the 
current European ships are of the order of several tens of kilometers.
\begin{table}
\centering
\small
\begin{tabular}[tb!]{|l|c|c|}
\hline 
 & Modele\,68 & Oto-Melara\tabularnewline
\hline 
\hline 
Caliber & 100\,mm & 76\,mm\tabularnewline
\hline 
Barrel length & 5.5\,m & 4.72\,m\tabularnewline
\hline 
Muzzle velocity & 870\,m/s & 925\,m/s\tabularnewline
\hline 
Weight of round & 23\,kg &12\,kg\tabularnewline
\hline
Weight of projectile & 13\,kg & 5--6\,kg\tabularnewline
\hline
Bursting-charge & 1\,kg & 0.4--0.75\,kg\tabularnewline
\hline
Rate of fire & 78\,rds/min & 80\,rds/min\tabularnewline
\hline 
Weight of turret & 22\,ton & 7.5\,ton\tabularnewline
\hline 
typical range & $<$17\,km & 20-30\,km\tabularnewline
\hline 
\end{tabular}
\vspace{2ex}\caption{\label{current_art}
Key parameters of current French and German standard naval
guns (data from \cite{navweaps_76mm, navweaps_100mm}).}
\end{table}

\section{Draft Railgun Dimensions}
To be able to calculate the electrical parameters of a railgun,
certain assumptions need to be made. From these assumptions a rough
draft for a future railgun system can be derived. This draft, in turn,
can be used to refine certain aspects and in an iterative process
improve the railgun definition. In this study only the first step is
done, resulting in a first sketch of a railgun. Calculations referred
to in \cite{mcnab_1} indicate that a projectile with a muzzle velocity 
of 2500\,m/s and a weight above 5\,kg will reach about 200\,nmi or
more. The velocity at the target will be of the order of 1000\,m/s to
1500\,m/s. Obviously the range and the final velocity is dependent on
the flight path, i.e. the fire angle, and on the aerodynamic
properties of the projectile. Nevertheless, without any further
studies, the above assumptions are not unrealistic and will be
discussed later in this paper. A 5\,kg projectile with a velocity at
the target of 1000\,m/s to 1500\,m/s velocity delivers a kinetic energy of
2.5\,MJ to 5.6\,MJ, resulting in approximately the same destructive energy
as conventional ammunition delivers explosively. For the acceleration
of the 5\,kg projectile in the railgun, an armature and sabot needs to
be added. The armature does supply the contact to the rails, while the
sabot mechanically attaches the projectile to the armature and acts as
a guide through the barrel. As an
estimate, an additional mass of 3\,kg is used to accommodate armature
and sabot. The total mass of the launch package is 8\,kg, resulting in
a muzzle energy of 25\,MJ. The muzzle energy of an electromagnetic
launcher can be calculated using:
\begin{equation}
E=\frac{1}{2}L'\, l\, I^{2} \label{eq_1},
\end{equation}
relating the inductance gradient $L'$, the acceleration length $l$
and the current $I$ to the energy. The inductance gradient $L'$
is to a large part determined by the geometry of the rails and the 
distance in between the rails (the caliber). For
practical, simple railguns with a square barrel 
a good first order approximation is $L'$~=~0.5\,$\mu$H/m.
Larger values can be obtained by using augmentation methods, adding
complexity and weight to the barrel of the launcher. To determine the
length of the barrel, a maximal allowed acceleration is assumed. For a 
constant acceleration the length is given by:
 \begin{equation}
 l=\frac{v^{2}}{2\, a}\label{eq_2}
 \end{equation}
Allowing an acceleration of 50\,kgee, the minimal length of the barrel
is 6.4\,m. Using this length and rearranging
formula (\ref{eq_1}) allows to determine the required current:
\begin{equation}
I=\sqrt{\frac{2\, E}{L'\, l}}\label{eq_3}
\end{equation}
With the values given, the current computes to $I=3.95$\,MA. This
current determines the minimal rail width from an electrical point of view. 
The maximum linear current density copper can sustain is approximately 
\mbox{$I'=I/(width\:of\:rails)\sim43$\,kA/mm}. This means, that the
minimal width is 92\,mm. To allow for a safety factor, a caliber of 
100\,mm is chosen.

\subsection{Electrical efficiency of the launcher}
\begin{figure}[tb!]
\centering
\includegraphics[width=3.5in]{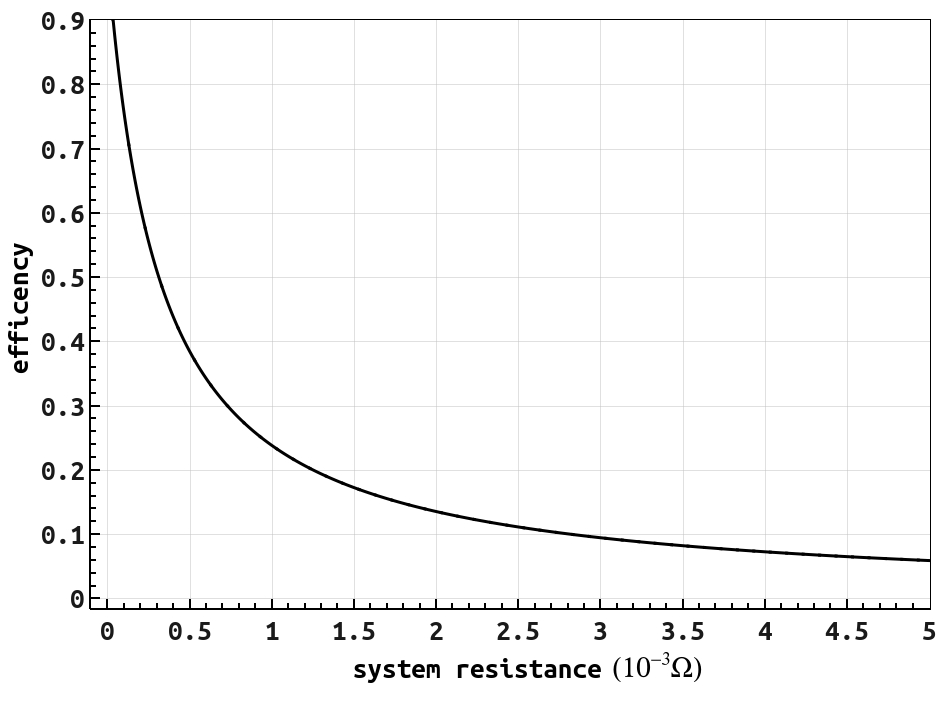}
\caption{System efficiency as a function of the total system resistance \cite{loeffler_1}.}
\label{sys_eff}
\end{figure}The amount of energy to be stored in the pulsed power system of the
railgun is determined by the muzzle energy multiplied by a factor
being inversely proportional to the efficiency (ratio of muzzle energy
to stored electrical energy) of the launch process. According to
\cite{loeffler_1}, the system efficiency is dependent on the inductance
gradient, on the projectile end-velocity and on the resistance of the
system. In this investigation, values for the inductance gradient and
the end-velocity were fixed (0.5\,$\mu$H/m and 2500\,m/s), leaving the resistance 
as the only parameter determining the overall system efficiency. Contributions to
this system resistance are: the power supply, the bus connecting the power
supply to the railgun, the rails and the contact resistance of the
armature. Figure \ref{sys_eff} shows the maximum system efficiency that
can be reached, given a certain value of resistance. This function
drops rapidly with an increasing resistance, reaching approx. 24\,\%
at 1\,m$\Omega$. The lower limit of the system resistance is the
contribution from the rails. For 50\,mm thick copper rails, with a
caliber of 100\,mm, this resistance calculates to 0.4\,m$\Omega$ at
full length and half (0.2\,m$\Omega$) of this value for the average 
value during a launch. Using this resistance as a guide, a total
system resistance of 0.5\,m$\Omega$ to 1\,m$\Omega$ is a realistic
assumption. From figure \ref{sys_eff}, this results in an efficiency of
24\,\% to 40\,\%. Using a value of 33\,\% for the overall efficiency, a
primary power supply unit being able to store 75\,MJ is required for
launcher operation. With a charger efficiency of 80\,\%, 1.6\,MW of
charging power is needed to allow one round per minute. For a 6 rounds
per minute operation one needs accordingly a charger being capable of
delivering 9.6\,MW of electrical power.

\section{Hypersonic Projectile}
\begin{figure}[t!]
\centering
\includegraphics[width=3.5in]{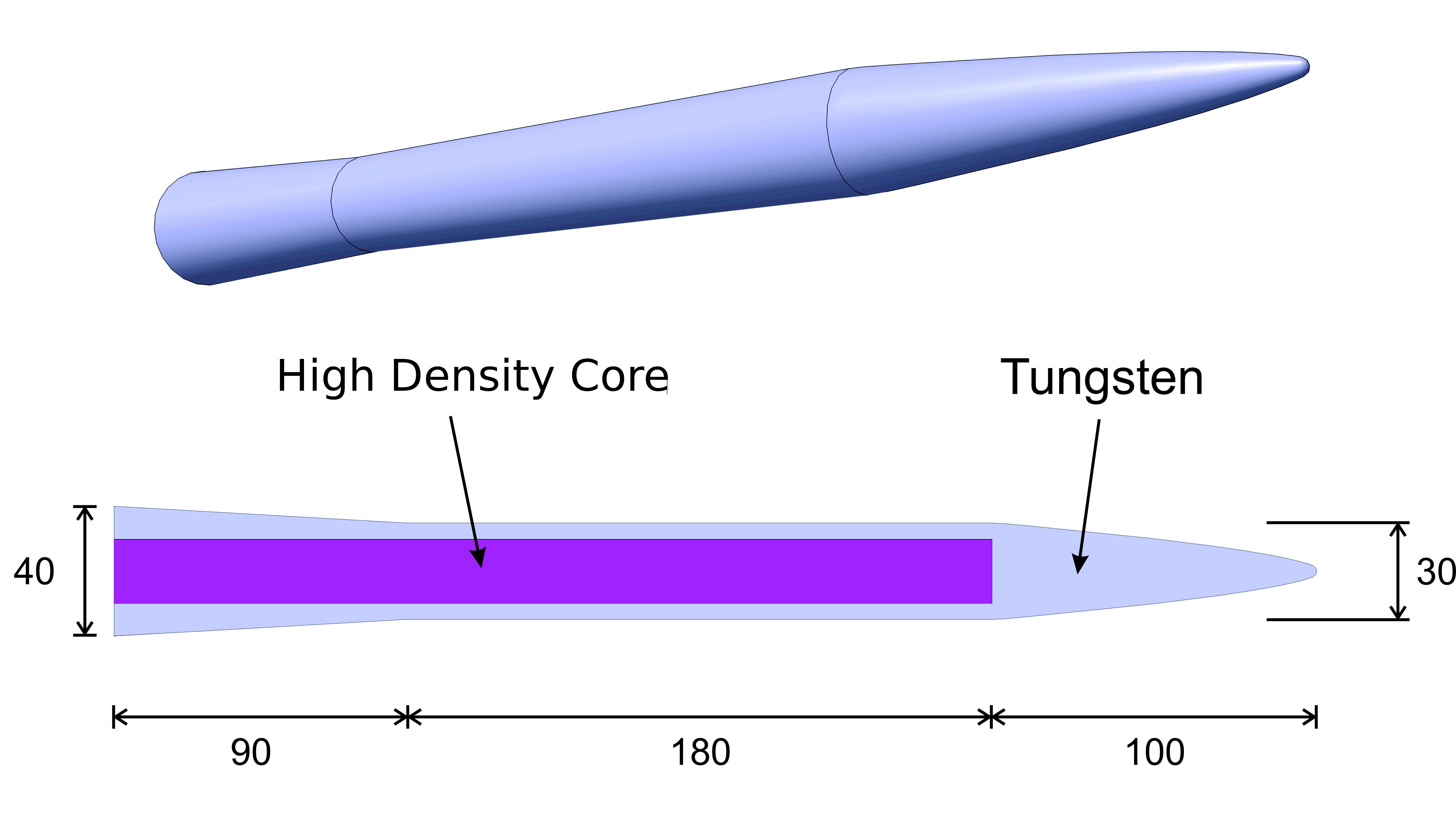}
\caption{Hypersonic kinetic energy projectile. All length are given in
millimeters.}
\label{projectile}
\end{figure}

The hypersonic kinetic energy projectile has to fulfill several requirements. To be
effective in the target, it shall transfer as much energy as possible to the target.
Therefore it needs to have a sufficiently high mass and a high
end-velocity. The large velocities experienced by the projectile
during its passage through the atmosphere require to pay special attention
during the design to low aerodynamic drag and heating. The expected surface temperature 
needs to be taken into account when choosing the projectile material.
Moreover the projectile needs to withstand the high acceleration forces.
For this application, the projectiles mass was chosen to be 5\,kg.
Tungsten was selected as material, because of its high density of
19\,g/cm$^3$ and high melting point of about 3420$^\circ$C. This material also 
increases the armor-piercing capabilities of the projectile. 
To further increase effectiveness, an additional high density core
material might be added. 
To reduce aerodynamic drag, a relatively small cross section of the projectile is
chosen, with the diameter of the projectile body being 30\,mm. The shape of the 
nose has a power-law form with a rounded nose-tip. This design is the best trade off 
between low aerodynamic drag and low aerodynamic heating. The total
length of the projectile is 370\,mm. For stable flight, the projectile
has a flare with a diameter of 40\,mm at the aft section, instead of fins.
Fins are more difficult to design in such a way that they can
withstand the expected high temperatures. Such a design would require a more detailed
and elaborate design study. A schematic drawing of the used
projectile geometry is shown in figure \ref{projectile}.

\section{Trajectory Simulation}
\begin{figure}[tb!]
\centering
\includegraphics[width=3.5in]{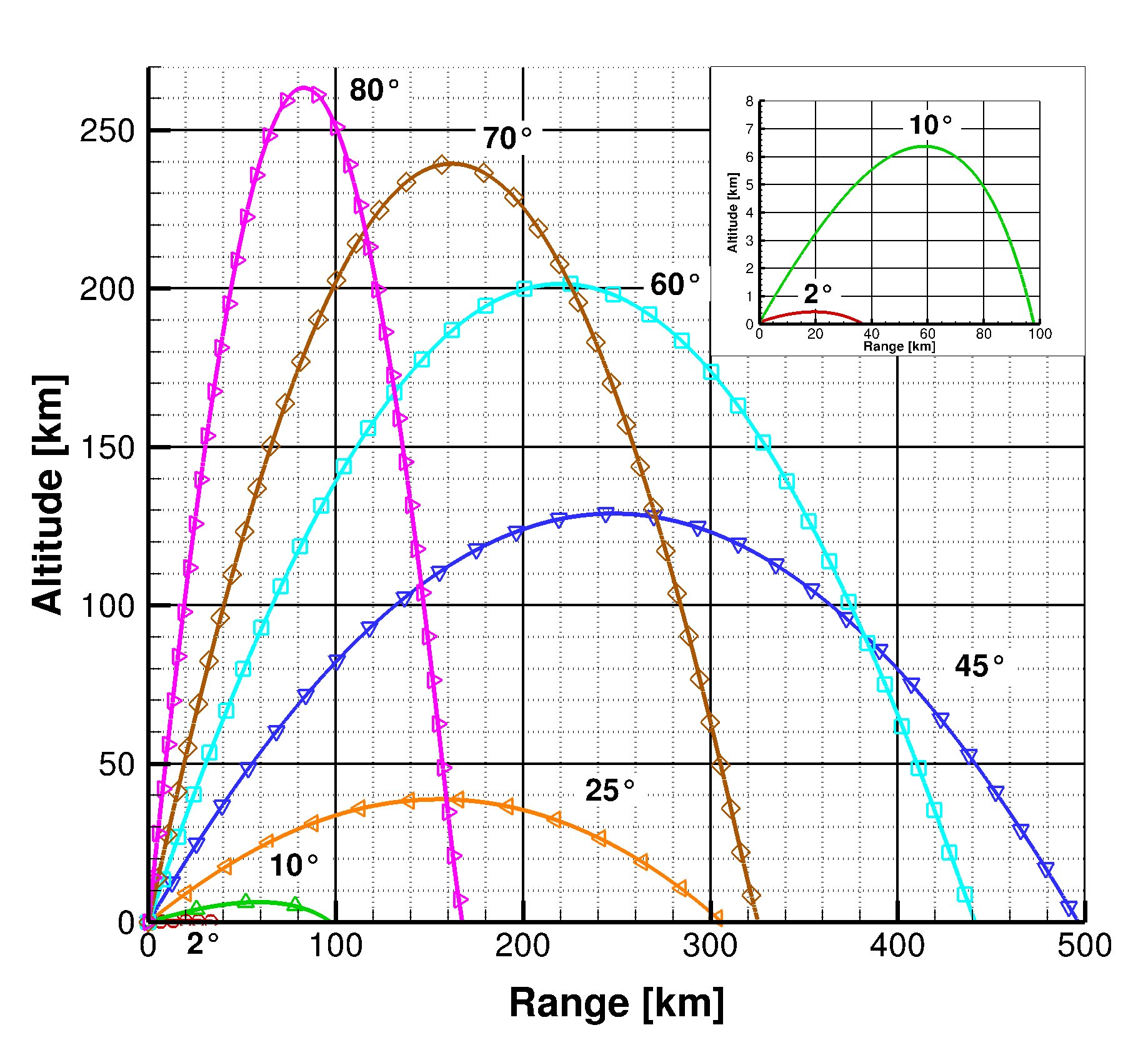}
\caption{Flight trajectories for the different launch angles.}
\label{trajectories}
\end{figure}

\begin{figure}[tb!]
\centering
\includegraphics[width=3.5in]{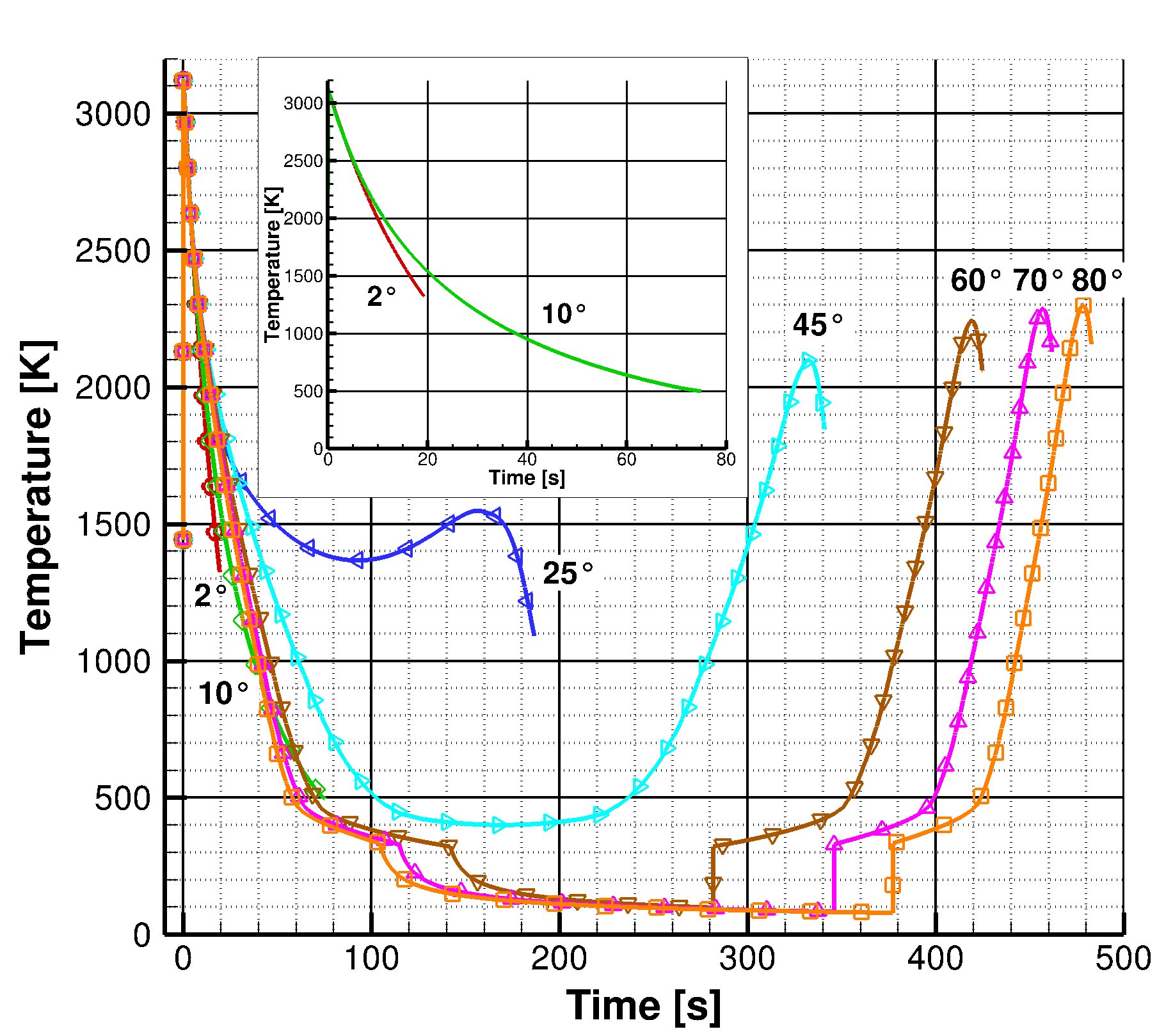}
\caption{Surface temperature at the stagnation point for the different
launching angles. The steps seen in the curves at around 300\,s are a 
result of the used atmospheric model, which sets the atmospheric 
density to zero above a height of 180 km.}
\label{temp}
\end{figure}

To calculate the flight path of the projectile, a coupled engineering tool is
used. The model is comprised out of a 6-DOF flight mechanics module, an
aerodynamics module, and a toolbox to calculate aerodynamic heating.
The flight mechanics module is an ordinary Runge-Kutta 4$^{th}$ order
solver for the equations of translation and rotation. The
aerodynamic part of the flight path is calculated using the industry
standard tool {\sc MISSLE DATCOM} \cite{missle_datcom}. The flight mechanics module
provides altitude, velocity, angle of attack and side slip angle for
each time step. {\sc MISSILE DATCOM} then derives the aerodynamics
coefficients. The aerodynamic forces and moments are calculated and
provided to the flight mechanics module. Aerodynamic heating is
calculated by means of the equation of Fay and Riddell \cite{stagnation}, to assess
the convective heat flux at the stagnation point. From the net heat
balance induced by convective heat flux and surface radiation, the
transient temperature distribution within the projectiles material
is determined using an implicit scheme for calculating the thermal
diffusion in a 1D-slice of the structure.
\begin{table}
\small
\centering
\begin{tabular}[tbh]{cccc}
\hline
Launch angle & v$_{hit}$ & E$_{hit}$ & Range \\
\hline\hline
2$^\circ$ &  1448\,m/s & 5.2\,MJ &  32\,km \\
10$^\circ$ &  637\,m/s &   1\,MJ & 100\,km \\
25$^\circ$ & 1270\,m/s &   4\,MJ & 303\,km \\
45$^\circ$ & 1791\,m/s &   8\,MJ & 496\,km \\
60$^\circ$ & 1919\,m/s & 9.2\,MJ & 420\,km \\
70$^\circ$ & 1958\,m/s & 9.6\,MJ & 311\,km \\
80$^\circ$ & 1975\,m/s & 9.8\,MJ & 132\,km \\
\hline
\end{tabular}
\vspace{2ex}
\caption{Results of the simulation for different launch angles. Shown
are the velocity and kinetic energy at the target v$_{hit}$ and
E$_{hit}$ and the horizontal distance between the launch and target
position.}
\label{sim_results}
\end{table}
The results for the simulation of the projectiles flight are shown in
table \ref{sim_results}. The launch angle was varied from 2$^\circ$ up to 80$^\circ$. 
The flight trajectories for the different angles are shown in figure
\ref{trajectories}. Depending
on the projectile launch angle the peak altitude can reach up to 260\,km 
height and the maximum range is about 500\,km for a launch angle of 45$^\circ$. 
From the different simulations, one can observe that a specific range
can be reached by two different launch angles, a flat and steep one
(as an example, see table \ref{sim_results}, the cases 25$^\circ$ and
70$^\circ$).
Using the steeper launch angle, a larger part of the trajectory goes through
space. Therefore, the distance the projectile has to pass through the
atmosphere, is reduced. This leads to an over the course of the flight 
reduced aerodynamic drag, resulting in a higher impact velocity. 
For the lower launch angle, the time of travel of the projectile is shorter, 
but the impact velocity and thus the impact energy are lower as well.
The spread is from an impact energy of about 1\,MJ for 10$^\circ$ up to
9.8\,MJ for a launching angle of 80$^\circ$. 
Figure \ref{temp} shows the temperature evolution at the stagnation point for
the different launch angles. As the large muzzle velocity results in a
large convective heat flux, the surface temperature is increasing
rapidly after launch to about 3100\,K. Soon, the velocity is decreasing 
and the altitude is increasing, both reducing the heating of the
projectile surface. Once above the atmosphere, the radiative cooling
allows for a further reduction in the temperature. Only when the
projectile reenters the atmosphere, the stagnation point temperature is
increasing again. At one point during the decent, the aerodynamic drag
in the increasingly dense atmosphere overcompensates the effect of gravity
and the projectile velocity decreases. This results in the turning
point in the temperature curves as seen for the launching angels above
25$^\circ$ at the very end of the flight.

\section{Summary}
\begin{table}
\small
\centering
\begin{tabular}[t!]{|l|c||l|c|}
\hline
Length          & 6.4\,m    & Prim. energy  & 75\,MJ\\ 
\hline
Caliber         & 100\,mm   & Muzzle energy & 25\,MJ\\
\hline
Projectile mass & 8\,kg     & Current       & 3.95\,MA\\
\hline
Muzzle velo.    & 2.5\,km/s & Acceleration  &$<$50\,kgee\\
\hline 
Range & up to 500\,km &Impact energy& up to 9.8\,MJ\\
\hline
\end{tabular}
\vspace{2ex}\caption{Key parameters for the shipboard railgun system}
\label{key_parameters}
\end{table}
Starting with a review of current, conventional marine artillery
systems, the key parameters of a first draft for a possible railgun
implementation were determined. The flight parameters of the
projectile were calculated using standard aerodynamic and flight
mechanic software. The results of this study are that a 
100\,mm square caliber railgun with a barrel length of 6.4\,m is able
to accelerate 8\,kg heavy launch packages. Depending on the launching
angle, the 5\,kg projectile will have a reach of up to 500\,km. For
this, the required primary electrical energy is of the order of
75\,MJ. Such a system would open up new ship artillery system capabilities. 
Further parameters of this gun are summarized in table \ref{key_parameters}.
It is the intention of the authors that the results of this study serve as 
a starting point for further discussions and studies about the capabilities 
and parameters of a future European shipboard railgun. 

\section*{Acknowledgment}
Part of this research was financed by the French Ministry of Defense
(DGA).

\end{document}